\documentclass{ws-ijmpa}

\begin{document}

\markboth{Zhi-zhong Xing and Shun Zhou} {The Minimal Type-I Seesaw
Model and Flavor-dependent Leptogenesis}

%
\catchline{}{}{}{}{}
%

\title{The Minimal Type-I Seesaw Model and Flavor-dependent Leptogenesis}

\author{Zhi-zhong Xing and
Shun Zhou$^*$}

\address{Institute of High Energy Physics, Chinese Academy of
Sciences, Beijing 100049, China \\$^*$E-mail:
zhoush@mail.ihep.ac.cn}

\maketitle

\begin{history}
\received{\today}
\end{history}

\begin{abstract}
In this talk, we first give a brief review of the so-called minimal
seesaw models and then concentrate on the minimal type-I seesaw
model with two almost degenerate right-handed Majorana neutrinos of
${\cal O}(1~{\rm TeV})$. A specific texture of the neutrino Yukawa
coupling matrix is proposed to achieve the nearly tri-bimaximal
neutrino mixing pattern. This ansatz predicts (1) $\theta^{}_{23} =
\pi/4$, $|\delta| = \pi/2$ and $\sin^2 \theta^{}_{12} = (1 - 2
\tan^2 \theta^{}_{13})/3$ in the $m^{}_1 = 0$ case; and (2)
$\theta^{}_{23} =\pi/4$ and $\theta^{}_{13} = \delta = 0$ in the
$m^{}_3 = 0$ case. In both cases, the cosmological baryon number
asymmetry can be explained via the resonant leptogenesis mechanism.
Finally, we demonstrate the significance of flavor-dependent effects
in our leptogenesis scenario.

\end{abstract}

\ccode{PACS numbers: 11.30.Fs, 14.60.Pq, 14.60.St}

\vspace{0.2cm}

Recent neutrino oscillation experiments have convinced us that
neutrinos are massive and lepton flavors are mixed.\cite{Strumia} In
order to accommodate tiny neutrino masses, one may immediately
extend the standard model by introducing three right-handed Majorana
neutrinos. The gauge-invariant Lagrangian relevant to lepton masses
reads
\begin{equation}
-{\cal L}^{}_{\rm mass} = Y^{}_l \overline{l^{}_{\rm L}} H E^{}_{\rm
R} + Y^{}_\nu \overline{l^{}_{\rm L}} \tilde{H} N^{}_{\rm R} +
\frac{1}{2} \overline{N^{c}_{\rm R}} M^{}_{\rm R} N^{}_{\rm R} +
{\rm h.c.} \; .
\end{equation}
After the spontaneous gauge symmetry breaking, the Dirac mass
matrices of charged leptons and neutrinos are given by $M^{}_l =
Y^{}_l v$ and $M^{}_{\rm D} = Y^{}_\nu v$, where $v = \langle H
\rangle \approx 174~{\rm GeV}$ is the vacuum expectation value of
the neutral Higgs field. The effective mass matrix of three light
neutrinos arises from the well-known seesaw formula $M^{}_\nu
\approx - M^{}_{\rm D} M^{-1}_{\rm R} M^T_{\rm D}$. The lightness of
left-handed Majorana neutrinos is therefore attributed to the
heaviness of right-handed Majorana neutrinos, while the phenomenon
of flavor mixing is due to the mismatch between the diagonalizations
of $M^{}_l$ and $M^{}_\nu$. One appealing advantage of the seesaw
models is the realization of baryogenesis via leptogenesis: the
lepton number asymmetries are first generated from the CP-violating
and out-of-equilibrium decays of heavy Majorana neutrinos and then
converted into the baryon number asymmetry by means of the
$(B-L)$-conserving sphaleron interaction.\cite{FY} However, a seesaw
model is usually plagued with too many free parameters even in the
basis where the mass matrices of charged leptons and right-handed
neutrinos are simultaneously diagonal. Let us explicitly count the
number of model parameters. Besides the right-handed Majorana
neutrino masses $M^{}_i$ (for $i=1, 2, 3$), there remain fifteen
free parameters in the Dirac neutrino mass matrix $M^{}_{\rm D}$
after three unphysical phases are removed by redefining the
charged-lepton fields. At low energies, there are only nine
observables: three light neutrino masses $m^{}_i$ (for $i = 1, 2,
3$), three neutrino mixing angles $\theta^{}_{ij}$ (for $ij = 12,
23, 13$), and three CP-violating phases $(\delta, \rho, \sigma)$.
Hence a generic seesaw model cannot make any specific predictions,
unless some additional assumptions are imposed on it.

The minimal type-I seesaw model includes only two heavy right-handed
Majorana neutrinos.\cite{MSM1,Rev} In this case, $M^{}_{\rm D}$ is a
$3 \times 2$ complex matrix which in general has nine physical
parameters. Hence there are totally eleven parameters in this
simplified seesaw model. In contrast, there are totally eighteen
parameters in the conventional seesaw model with three right-handed
Majorana neutrinos. An intrinsic feature of the minimal type-I
seesaw model is that the lightest neutrino must be massless and only
one of the Majorana CP-violating phases ($\rho$ or $\sigma$)
survives. It is well known that this is the most economical seesaw
scenario that can interpret both the neutrino masses and the
cosmological baryon number asymmetry. Alternatively, the
introduction of a heavy triplet scalar into the standard model can
also give rise to tiny Majorana masses of three known
neutrinos.\cite{Valle} From the viewpoint of grand unified theories,
however, the most natural choice is to introduce both the
right-handed neutrino singlets and the scalar triplet. It is easy to
show that only one triplet Higgs and one right-handed Majorana
neutrino are enough to account for both the neutrino oscillation
experiments and the observed matter-antimatter asymmetry of the
universe. This scenario is usually referred to as the minimal
type-II seesaw model.\cite{MSM2}

We now propose an intriguing minimal type-I seesaw model with two
nearly degenerate right-handed Majorana neutrinos of ${\cal O}(1 ~
\rm TeV)$, which may be accessible at the forthcoming Large Hadron
Collider. Let us begin with a useful parametrization:\cite{XZ}
\begin{equation}
M^{(1)}_{\rm D} \; = \; V^{}_0 \left(\begin{matrix} 0 & 0 \cr x & 0
\cr 0 & y \cr \end{matrix} \right) U \;, ~~~~~ M^{(3)}_{\rm D} \; =
\; V^{}_0 \left(\begin{matrix} x & 0 \cr 0 & y \cr 0 & 0 \cr
\end{matrix} \right) U \;
\end{equation}
for $m^{}_1 = 0$ and $m^{}_3 = 0$ cases, where $V^{}_0$ and $U$ are
$3 \times 3$ and $2 \times 2$ unitary matrices, respectively. Then
the seesaw relation $M^{}_\nu = - M^{}_{\rm D} M^{-1}_{\rm R}
M^T_{\rm D}$ implies that the neutrino mixing depends primarily on
$V^{}_0$ and the decays of heavy neutrinos rely mainly on $U$. Hence
we take $V^{}_0$ to be the tri-bimaximal mixing pattern\cite{TB}
\begin{equation}
V^{}_0 \; = \; \left( \begin{matrix} 2/\sqrt{6} & 1/\sqrt{3} & 0 \cr
-1/\sqrt{6} & 1/\sqrt{3} & 1/\sqrt{2} \cr 1/\sqrt{6} & ~~
-1/\sqrt{3} ~~ & 1/\sqrt{2} \cr \end{matrix} \right) \; ,
\end{equation}
which is compatible very well with the best fit of current
experimental data.\cite{Strumia} On the other hand, the unitary
matrix $U$ can be parameterized as
\begin{equation}
U \; = \; \left ( \begin{matrix} \cos \vartheta & \sin \vartheta \cr
-\sin \vartheta & \cos \vartheta \cr \end{matrix} \right ) \left (
\begin{matrix} e^{-i\alpha} & 0 \cr 0 & e^{+i\alpha} \cr \end{matrix} \right) \; .
\end{equation}
Since $\alpha$ is the only phase parameter in our model, it should
be responsible both for the CP violation in neutrino oscillations
and for the CP violation in $N^{}_i$ decays. For simplicity, here we
fix $\vartheta = \pi/4$ and highlight the role of $\alpha$ in
neutrino mixing and leptogenesis. In order to implement the idea of
resonant leptogenesis, we suppose that two heavy Majorana neutrino
masses are highly degenerate; i.e., the magnitude of $r \equiv
(M^{}_2 - M^{}_1)/M^{}_2$ is strongly suppressed.\cite{PU}

Given $|r| < {\cal O}(10^{-4})$, the explicit form of $M^{}_\nu$ can
reliably be formulated from the seesaw relation $M^{}_\nu = -
M^{}_{\rm D} M^{-1}_{\rm R} M^T_{\rm D}$ by neglecting the tiny mass
splitting between $N^{}_1$ and $N^{}_2$. In this good approximation,
we diagonalize $M^{}_\nu$ through $V^\dagger M^{}_\nu V^* = {\rm
Diag} \left \{ m^{}_1, m^{}_2, m^{}_3 \right \}$, where $V$ is just
the neutrino mixing matrix. In the $m^{}_1 = 0$ case, three mixing
angles and the Dirac CP-violating phase are determined by
\begin{equation}
\sin^2 \theta^{}_{12} = \frac{1 - \sin^2 \theta}{3 - \sin^2 \theta}
\; , ~~ \theta^{}_{23} = \frac{\pi}{4} \; , ~~ \sin^2 \theta^{}_{13}
= \frac{\sin^2 \theta}{3} \; ,  ~~\delta = - \frac{\pi}{2} \; ,
\end{equation}
where $\theta$ is given by $\tan 2\theta = 2 \omega \tan 2\alpha
/(1+\omega^2)$ with $\omega \equiv x/y \in (0, 1)$. It is
straightforward to derive $\sin^2 \theta^{}_{12} = (1 - 2 \tan^2
\theta^{}_{13})/3$. Taking account of $m^{}_2 = \sqrt{\Delta
m^2_{21}}$ and $m^{}_3 = \sqrt{\Delta m^2_{21} + |\Delta
m^2_{32}|}$, we arrive at $m^{}_2 \approx 8.9 \times 10^{-3}$ eV and
$m^{}_3 \approx 5.1 \times 10^{-2}$ eV by using $\Delta m^2_{21}
\approx 8.0 \times 10^{-5} ~ {\rm eV}^2$ and $|\Delta m^2_{32}|
\approx 2.5 \times 10^{-3} ~ {\rm eV}^2$ as the typical
input.\cite{Strumia} The values of $m^{}_2$ and $m^{}_3$, together
with $\theta^{}_{13} < 10^\circ$, lead to $0.39 \lesssim \omega
\lesssim 0.42$, $0^\circ \lesssim \alpha \lesssim 23^\circ$ and
$0^\circ \lesssim \theta \lesssim 18^\circ$. We find that $m^{}_2
\approx x^2/M^{}_2$ and $m^{}_3 \approx y^2/M^{}_2$ are good
approximations for $\alpha \lesssim 10^\circ$. For the inverted
neutrino mass hierarchy ($m^{}_3 = 0$), we have
\begin{eqnarray}
\sin^2 \theta^{}_{12} = \frac{1 + \sin^2 \theta}{3} \; , ~~~~
\theta^{}_{23} = \frac{\pi}{4}  \; , ~~~~ \theta^{}_{13} = 0\; ,
~~~~ \delta = 0 \; .
\end{eqnarray}
Taking account of $m^{}_1 = \sqrt{|\Delta m^2_{32}| - \Delta
m^2_{21}}$ and $m^{}_2 = \sqrt{|\Delta m^2_{32}|}$, we get $m^{}_1
\approx 4.9 \times 10^{-2}$ eV and $m^{}_2 \approx 5.0 \times
10^{-2}$ eV by inputting $\Delta m^2_{21} \approx 8.0 \times 10^{-5}
~ {\rm eV}^2$ and $|\Delta m^2_{32}| \approx 2.5 \times 10^{-3} ~
{\rm eV}^2$. Given $30^\circ < \theta^{}_{12} <
38^\circ$,\cite{Strumia} $\theta$ is found to lie in the range $0
\lesssim \theta \lesssim 22^\circ$. Furthermore, the values of
$m^{}_1$ and $m^{}_2$ allow us to get $0^\circ \lesssim \alpha
\lesssim 22^\circ$ and $0.991 \lesssim \omega \lesssim 0.992$. The
neutrino masses reliably approximate to $m^{}_1 \approx x^2/M^{}_2$
and $m^{}_2 \approx y^2/M^{}_2$ for $\alpha \lesssim 10^\circ$ in
this case. From a phenomenological point of view, the scenario with
$m^{}_1 =0$ is more favored and more interesting than the scenario
with $m^{}_3 =0$. Both of them can be tested in the near future.

We proceed to calculate the cosmological baryon number asymmetry via
the flavor-dependent resonant leptogenesis. Note that all the Yukawa
interactions of charged leptons are in thermal equilibrium at the
TeV scale, so the lepton number asymmetry for each flavor should be
treated separately due to its distinct thermal history.\cite{Flavor}
In the framework of resonant leptogenesis, it is straightforward to
compute the CP-violating asymmetry between $N^{}_i \rightarrow
l^{}_\alpha + H^{\rm c}$ and $N^{}_i \rightarrow l^{\rm c}_\alpha +
H$ decays for each lepton flavor $\alpha$ ($=e$, $\mu$ or $\tau$):
\begin{equation}
\varepsilon^{}_{i \alpha} = \frac{8\pi \left(M^2_i - M^2_j\right)
{\rm Im}\left\{ \left(Y^{}_\nu\right)^{}_{\alpha j} \left(Y^{}_{\nu
}\right)^{*}_{\alpha i} M^{}_i \left[M^{}_i \left(Y^\dagger_\nu
Y^{}_\nu \right)^{}_{ji} + M^{}_j \left(Y^\dagger_\nu Y^{}_\nu
\right)^{}_{ij} \right] \right\}}{\left[ 64\pi^2 \left(M^2_i
-M^2_j\right)^2 + M^4_i \left(Y^\dagger_\nu Y_\nu \right)^{2}_{jj}
\right]\left(Y^\dagger_\nu Y_\nu \right)^{}_{ii}} \; ,
\end{equation}
where $i$ and $j$ run over $1$ and $2$ but $i \neq j$. To take
account of the inverse decays and lepton-number-violating scattering
processes, we define the corresponding decay parameters $K^{}_{i
\alpha} \equiv \left| (Y^{}_\nu)^{}_{\alpha i} \right|^2
K^{}_i/\left(Y^\dagger_\nu Y_\nu \right)^{}_{ii}$, where $K^{}_i
\equiv \Gamma^{}_i/H$ at $T=M^{}_i$ with $\Gamma^{}_i =
\left(Y^\dagger_\nu Y_\nu \right)^{}_{ii}M^{}_i/(8\pi)$ being the
total decay width of $N^{}_i$ and $H(T) = 1.66 \sqrt{g^{}_*}
T^2/M^{}_{\rm planck}$ being the Hubble parameter. Note that these
quantities can be explicitly figured out with the help of Eqs. (2),
(3) and (4) as well as the constraints from neutrino masses and
mixing parameters. In the strong washout regime $K^{}_{i\alpha}
\gtrsim 1$, the survival of lepton number asymmetries is
approximately characterized by the efficiency factors\cite{Bari}
\begin{eqnarray}
\kappa^{}_{i \alpha} \approx \frac{2}{K^{}_\alpha z^{}_{\rm
B}(K^{}_\alpha)}\left[1-\exp\left(-\frac{K^{}_\alpha z^{}_{\rm
B}(K^{}_\alpha)}{2}\right)\right] \; ,
\end{eqnarray}
with $z^{}_{\rm B}(K^{}_\alpha) \simeq 2+4 K^{0.13}_{\alpha}
\exp\left(-2.5/K^{}_\alpha\right)$ and $K^{}_\alpha = \sum_i K^{}_{i
\alpha}$. The final baryon number asymmetry can be estimated as
$\eta^{\rm f}_{\rm B} \approx -0.96 \times 10^{-2} \sum_i
\sum_\alpha \left (\varepsilon^{}_{i \alpha} \kappa^{}_{i \alpha}
\right)$. It is found that the observed baryon number asymmetry
$\eta \approx 6.1 \times 10^{-10}$ can indeed be achieved.\cite{XZ}
For comparison, we also examine the final baryon asymmetry in the
one-flavor approximation. Summing the CP-violating asymmetries over
all flavors $\varepsilon^{}_i = \sum_\alpha \varepsilon^{}_{i
\alpha}$ and using the flavor-independent efficiency factors
$\kappa^{}_i \approx \; 0.5/\left( \sum_i K^{}_i \right)^{1.2}$, one
gets the final baryon asymmetry $\eta^{}_{\rm B} \approx -0.96
\times 10^{-2} \sum_i \left (\varepsilon^{}_{i} \kappa^{}_{i}
\right)$. In our scenario,
\begin{eqnarray}
\frac{\eta^{\rm f}_{\rm B}}{\eta^{}_{\rm B}} = \frac{\sum_{i,
\alpha} \varepsilon^{}_{i \alpha} \kappa^{}_{i \alpha}}{\sum_i
\varepsilon_i \kappa^{}_i} \approx \left \{ \begin{matrix}-0.71 ~~~~
(m^{}_1 =0) \; , \cr +52.1 ~~~~ (m^{}_3 =0) \; . \cr \end{matrix}
\right . \;
\end{eqnarray}
We see that the flavor-independent prediction for $\eta^{}_{\rm B}$
becomes negative in the $m^{}_1 = 0$ case, while it is enhanced by a
factor $\sim 50$ in the $m^{}_3 = 0$ case.

\end{document}